\documentclass{ws-ijmpd}
\usepackage{url}
\usepackage[super,compress]{cite}
\usepackage{tensor}
\usepackage[colorlinks=true,linkcolor=blue,citecolor=blue,urlcolor=blue]{hyperref}
\newcommand{\eqr}{\eqref}
\newcommand{\Q}{\mathcal{Q}}
\newcommand{\dr}{\ell}
\newcommand{\g}{g\indices}

\begin{document}

\markboth{L.~Rodriguez, S.~Rodriguez, R.~Suleiman, et~al.}
{NC, Entropy Universality and CFT Duality in CWG}

%
\catchline{}{}{}{}{}
%

\title{N\"other Currents, Black Hole Entropy Universality and CFT Duality in Conformal Weyl Gravity}
\author{Daksh Aggarwal$^1$, Dominic Chang$^{2,3}$, Quentin Dancewicz Helmers$^4$, Nesibe Sivrioglu$^4$, L. R. Ram-Mohan$^5$, Leo Rodriguez$^{4,5}$\footnote{rodriguezl@grinnell.edu}, Shanshan Rodriguez$^4$\footnote{rodriguezs@grinnell.edu} and Raid Suleiman$^{3,5}$\footnote{rsuleiman@cfa.harvard.edu}}

\address{$^1$Department of Mathematics, Brown University\\Providence RI, 02912, USA\\
$^2$ Black Hole Initiative, Harvard University\\Cambridge, MA 02138, USA\\
$^3$ Center for Astrophysics, Harvard \& Smithsonian\\Cambridge MA, 02138, USA\\
$^4$ Department of Physics, Grinnell College\\Grinnell IA, 50112, USA\\
$^5$ Department of Physics, Worcester Polytechnic Institute\\Worcester MA, 01609, USA}
\maketitle

\begin{history}
\received{Day Month Year}
\revised{Day Month Year}
\end{history}

\begin{abstract}
In this paper we study black hole entropy universality within the Conformal Weyl gravity paradigm. We do this by first computing the entropy of specific vacuum and non-vacuum solutions, previously unexplored in Conformal Weyl gravity via both the N\"other current method and Wald's entropy formula. For the vacuum case, we explore the near horizon near extremal Kerr metric, which is also a vacuum solution to Conformal Weyl gravity and not previously studied in this setting. For the non-vacuum case we couple the conformal Weyl gravity field equations to a near horizon (linear) $U(1)$ gauge potential and analyze the respective found solutions. We highlight the non-universality of black hole entropy between our studied black hole solutions of varying symmetries. However despite non-universality, the respective black hole entropies are in congruence with Wald's entropy formula for the specific gravity theory. Finally and despite non-universality, we comment on the construction of a near horizon CFT dual to one of our unique non-vacuum solutions. Due to the non-universality, we must introduce a parameter (similarly to entropy calculations in LQG) which we also call $\gamma$ and relating to the Weyl anomaly coefficient. The  construction follows an $AdS_2/CFT_1$ correspondence in the near horizon, which  enables the computation of the full asymptotic symmetry group of the chosen non-vacuum conformal Weyl black hole and its near horizon quantum CFT dual. We conclude with a discussion and outlook for future work.
\end{abstract}




\section{Introduction}
Conformal Weyl gravity (CWG) and fourth order gravity theories relating to string theory and in general \cite{Brensinger:2017gtb,Brensinger:2020gcv} have recently reemerged as helpful tools in gravitational/cosmological physics \cite{Brensinger:2019mnx,Mannheim:2009qi,Edery:1997hu,Edery:1998zi,Edery:2001at}, black hole physics \cite{Zou:2020rlv,Momennia:2019cfd,Momennia:2019edt,Momennia:2018hsm}, their thermodynamics \cite{Xu:2018liy,Bambi:2017ott} and role in quantum gravity \cite{Mannheim:2009qi,Stelle:1976gc,Edery:2006hg}. CWG is given by:
\begin{align}\label{eq:CWGA}
S_{CWG}=\alpha_{c}\int d^4x\sqrt{-g}W^{\alpha\mu\beta\nu}W_{\alpha\mu\beta\nu},
\end{align}
which is a purely covariant and conformally invariant Weyl tensor squared action \cite{Mannheim:2011ds,Kiefer:2017nmo} for unit-less coupling $\alpha_{c}$. In addition, CWG includes all of the Einstein-Hilbert (EH) vacuum solutions as part of its solution space and cosmological dynamics appear naturally within this formalism \cite{Mannheim:1988dj,Mannheim:2005bfa}. Additionally, several Einstein-Hilbert non-vacuum solutions have analogue counterparts within CWG \cite{Mannheim:1990ya}. The above leads to specific regimes where the two theories may be considered equivalent \cite{Maldacena:2011mk,Anastasiou:2016jix,Anastasiou:2020mik}.

However, the fourth order nature and unitarity issues endow it with some difficulties. It is nonetheless an attractive theory and, since its vacuum solution space includes Einstein-Hilbert vacua, CWG includes all the standard solar-system tests of general relativity and warrants investigation on its own right. One aspect and line of questioning that has not been extensively explored yet is the implementation of the near horizon black hole conformal field theory duality paradigm (with the $Kerr/CFT$ correspondence \cite{kerrcft,Compere:2012jk} as a notable example) within CWG. Since the seminal work of Brown and Henneaux \cite{brownhenau} and the $Kerr/CFT$ correspondence, it has become universally accepted that black holes are holographically dual to lower dimensional conformal field theories ($CFT$) at their boundaries. This is reminiscent of the more general anti-de Sitter($AdS$)/$CFT$ correspondence of string theory \cite{Maldacena:1997re}. This paradigm has surged a tremendous program led by Strominger \cite{strom2,kerrcft}, Carlip \cite{Carlip:2011ax,carlip,carlip3,carlip2}, Park \cite{Park:1999tj,Park:2001zn,kkp} and others, in applying $CFT$ techniques to compute Bekenstein-Hawking entropy of various black holes, in particular (near-)extremal ones. This is based upon the feature that (near-)extremal black holes admit a local $AdS_2\times S^2$ topology in their near horizons. This fact allows for the implementation of $AdS/CFT$ techniques to construct lower dimensional black hole $CFT$ duals, effectively providing a statistical quantum mechanical description of black hole horizon microstates. The results of these dualities include the standard black hole thermodynamic quantities \cite{Button:2010kg,Button:2013rfa,hawk2,hawk3,beken},
\begin{align}
\label{eq:htaen}
\begin{cases}
T_H=\frac{\hbar\kappa}{2\pi}&\text{Hawking Temperature}\\
S_{BH}=\frac{A}{4\hbar G}&\text{Bekenstein-Hawking Entropy}
\end{cases},
\end{align}
which are the usual tests for any candidate theory of quantum gravity. 

In general, the argument is that the asymptotic symmetry group (ASG), which preserves specific metric fall off conditions, is given by a Virasoro algebra with a calculable center:
\begin{align}
\label{eq:vir}
\left[\Q_m,\Q_n\right]=(m-n)\Q_{m+n}+\frac{c}{12}m\left(m^2-1\right)\delta_{m+n,0},
\end{align}
where $m,n\in\mathbb{Z}$. Then using Cardy's formula \cite{cardy2,cardy1}:
\begin{align}
\label{eq:cf}
S=2\pi\sqrt{\frac{c\cdot\Q_0}{6}}.
\end{align}
to count the horizon microstates, in terms of the dual $CFT$'s central charge $c$ and normalized lowest eigenmode $\Q_0$, provides a holographic computation of black hole entropy.
Traditionally, a finite time regulator in terms of surface gravity is employed to regulate the quantum charges of \eqr{eq:vir}. This leads to difficulty in using \eqr{eq:cf} for extremal black holes, where surface gravity, and thus Hawking temperature, vanishes identically. To circumvent this, a thermal Cardy formula may be used: 
\begin{align}
\label{eq:tcf}
S\sim\left(cT\right)
\end{align}
where $T$ is the inverse finite time regulator which for extremal black holes corresponds to the sum of left and right Frolov-Thorne vacuum temperatures for chiral $CFT$s. Additionally,  $c=c_L=c_R$ assuming diffeomorphism invariance of the two chiral halves \cite{Kraus:2005zm,Compere:2012jk,Castro:2010fd,cft}.

There has been recent and semi-recent interest in black hole holography applied within CWG and other fourth order gravity theories in general \cite{Liu:2021hvb,Rodriguez:2017nxe,Li:2015mda,Grumiller:2013mxa,Krishnan:2009tj}, with several interesting results, some of which that do not align with expectations from general relativity. One such misalignment is the loss of universality, i.e., black hole entropy still exhibits an area law in CWG, but the coefficient is not always $1/4$ and in fact differs depending on what spacetime is chosen. This loss of universality can immediately be deduced by inspection of the the Wald entropy for CWG \cite{Iyer:1994ys,Iyer:1995kg,Lu:2012xu,Li:2015mda}:
\begin{align}\label{eq:CWGWEnt}
S=-\frac{\alpha_c}{8}\int W_{\mu\nu\alpha\beta}\epsilon^{\mu\nu}\epsilon^{\alpha\beta}d\Sigma,
\end{align}
where $d\Sigma$ is the orthogonal hypersurface area element and $\epsilon^{\mu\nu}$ is its unit normal bivector. From the above we see that the Wald entropy in CWG depends on the Weyl tensor. Thus black holes that are conformally flat, with universal area entropy laws in general relativity, will exhibit zero entropy in CWG. Specific examples include the $RN/CFT$ correspondence \cite{Garousi:2009zx,Chen:2010bsa,Chen:2011gz,Chen:2009ht} with spacetime element:
\begin{align}
ds^2=Q^2\left(-r^2dt^2+\frac{dr^2}{r^2}+d\Omega^2\right),
\end{align}
where $Q$ is the black hole charge parameter and $d\Omega^2$ is the unit two sphere. The above line element is conformally flat for the spacetime diffeomorphism and conformal factor ($e^{2\omega}$) given by:
\begin{align}
T=t,~x=-\frac{\sin\theta\cos\phi}{r},~y=-\frac{\sin\theta\sin\phi}{r},~z=-\frac{\cos\theta}{r},~\text{and}~e^{2\omega}=Q^2r^2,
\end{align}
and thus has vanishing Weyl tensor. Another example is the Weyl rescaled near horizon Schwarzschild spacetime in finite mass/temperature gauge \cite{Rodriguez:2017nxe,cadss}:
\begin{align}
ds^2=-\frac{r^2-2GMr}{\dr^2}dt^2+\frac{\dr^2}{r^2-2GMr}dr^2+{\dr^2}d\theta^2+{\dr^2}\sin^2\theta d\phi^2,
\end{align}
which is conformally flat via the diffeomorphism and conformal factor:
\begin{align}
\begin{split}
T=&\frac{\exp{\left\{\frac{t}{4GM}\right\}}}{GM}\cosh u;~x=\frac{\exp{\left\{\frac{t}{4GM}\right\}}}{GM}\sinh u\cos\phi\sin\theta;\\
y=&\frac{\exp{\left\{\frac{t}{4GM}\right\}}}{GM}\sinh u\sin\phi\sin\theta;~z=\frac{\exp{\left\{\frac{t}{4GM}\right\}}}{GM}\sinh u\cos\theta;\\
e^{2\omega}=&\exp{\left\{\frac{-t}{2GM}\right\}}\ell^2r(r-2GM);\text{and}~u=\ln\sqrt{1-\frac{2GM}{r}}.
\end{split}
\end{align}
The last black hole example we discuss here was first introduced in \cite{Rodriguez:2017nxe} by directly forging for a local $AdS_2\times S^2$ CWG vacuum solution with the Kaluza-Klein ansatz:
\begin{align}\label{eq:KKSolAnz}
ds^2=&K_1(\theta)\left[-f(r)dt^2+f(r)^{-1}dr^2+\dr^2d\theta^2\right]+\frac{\dr^2sin^2{\theta}}{K_2(\theta)} \left(d\phi-A_\mu(r)dx^\mu\right)^2
\end{align}
and yielding:
\begin{align}\label{eq:my1stsol}
\begin{split}
ds^2=&\sin{\theta}\left[-\frac{r^2-2GMr}{\ell^2}dt^2+\frac{\ell^2}{r^2-2GMr}dr^2+\dr^2d\theta^2\right]\\
&+\dr^2\sin{\theta} \left(d\phi+\frac{r-2GM}{\dr^2}dt\right)^2.
\end{split}
\end{align}
This spacetime is very interesting since it is a vacuum solution with local axisymmetry and $AdS_2\times S^2$ topology with respect to CWG, however from a general relativistic interpretation it is globally of \emph{Petrov Type O} and \emph{Segre Type} $[(111),1]$\footnote{We employ the standard spacetime classification notation of Ref.~\refcite{Stephani:2003tm}.}. In other words (and not previously discussed in Ref.~\refcite{Rodriguez:2017nxe}), from an Einstein perspective it is conformally flat and sourced by a perfect fluid of energy momentum tensor:
\begin{align}
\begin{split}
T_{\mu\nu}=&(\rho+p)U_\mu U_\nu+p g_{\mu\nu},~\text{where}\\
\rho=&p=-\frac{3\csc^3\theta}{32\pi G\ell^2}~\text{and}~U^\mu=\left(0,0,\frac{i\sqrt{\csc\theta}}{\ell},0\right).
\end{split}
\end{align}
The diffeomorphism and conformal factor which maps \eqref{eq:my1stsol} to conformal flatness is given by:
\begin{align}
\begin{split}
T=&2e^{\alpha}\sqrt{\frac{r}{r-2GM}}\cosh \frac{\phi}{2},~x=2e^{\alpha}\sqrt{\frac{r}{r-2GM}}\sinh \frac{\phi}{2},\\
y=&2e^{\alpha}\sqrt{\frac{2GM}{r-2GM}}\sin \frac{\theta}{2},~z=2e^{\alpha}\sqrt{\frac{2GM}{r-2GM}}\cos \frac{\theta}{2},\\
e^{2\omega}=&(r-2GM)\exp{\left\{-\frac{t}{\ell}+\frac{\ell}{2GM}\phi\right\}}\sin\theta,\text{and}~\alpha=\frac{t}{2\ell}-\frac{\ell}{4GM}\phi.
\end{split}
\end{align}

As mentioned earlier, there has been recent interest in black hole holography and CWG. Specifically, and for a non-zero result, the entropy for spherical black holes:
\begin{align}\label{eq:SphSymBH}
ds^2=-f(r)dt^2+\frac{1}{f(r)}+r^2d\Omega^2
\end{align}
that are CWG vacuum solutions and not conformally flat were studied from the holographic perspective via the N\"other current method \cite{Majhi:2012nq,Majhi:2011ws,Majhi:2012tf,Ashworth:1998uj} in Ref.~\refcite{Li:2015mda} and shown to agree with the Wald entropy \eqref{eq:CWGWEnt}. In particular, for this case the result obtained was:
\begin{align}\label{eq:LieEntRes}
S=\frac{\alpha_c}{8}\left(\frac{4}{3r^2_+}+\frac{4f'\left(r_+\right)}{3r_+}-\frac{2f''\left(r_+\right)}{3}\right)A,
\end{align}
where $r_+$ is the horizon radius of \eqref{eq:SphSymBH} and $A$ its horizon area. Our goal will be to add to the above examples by studying the holographic $CFT$ duals of some additional black holes that arise within the solution space of CWG. 

We do this by first providing a quick review of the boundary N\"other current method in Sec.~\ref{sec:BNCRev}. This method has been extensively explored for spherically symmetric vacuum black holes in their respective gravity theories. As a new and previously unexplored example of how the method is used to compute black hole entropy in general relativity, we first apply it to the rotating Kerr spacetime. This calculation results in the standard Bekenstein-Hawking entropy of the Kerr black hole, exemplifying entropy universality of general relativity. In Sec.~\ref{sec:CQMNHNEK} we, for the firs time, employ and extend the N\"other current method in CWG to the Kerr spacetime, which is also a vacuum solution to CWG. We compute its entropy using this method which aligns with the respective Wald entropy \eqref{eq:CWGWEnt}, but differs from the area law in \eqref{eq:LieEntRes}. In Sec.~\ref{sec:CQM-Max} we focus on near horizon non-vacuum CWG solutions. To do so, we first couple the CWG field equations to a linear $U(1)$ potential and forge for solutions with local $AdS_2\times S^2$ topology. We then compute black hole entropy by extension of the N\"other current method and compare to the respective Wald entropy \eqref{eq:CWGWEnt}. The entropy results again differ from \eqref{eq:LieEntRes} and form each other. Finally, in Sec.~\ref{sec:QFieldsICST} and in light of non-universality, we comment on the possible construction of suitable two dimensional $CFT$s that would be holographically dual to our chosen black hole examples of CWG within an $AdS_2/CFT_1$ correspondence. We end our study with a discussion of results and outlook for future directions. 
\section{Quick Review of the Boundary N\"other Current and Its Charge}\label{sec:BNCRev}
In this part we will briefly review the boundary N\"other current method according to Ref.~\refcite{Majhi:2012nq,Majhi:2011ws,Majhi:2012tf,Ashworth:1998uj}. We will begin with the boundary term of a general Lagrangian density $\mathcal{L}$, written as a total divergence, such that:
\begin{align}\label{eq:genBact}
\sqrt{g}\mathcal{L}=\sqrt{g}\nabla_\mu B^\mu.
\end{align}
Next, considering an infinitesimal diffeomorphism $x^\mu\to x^\mu+\xi^\mu$ applied to the left hand side of \eqref{eq:genBact} we have:
\begin{align}\label{eq:Lhsga}
\delta_\xi\left(\sqrt{g}\mathcal{L}\right)=\delta_\xi\sqrt{g}\mathcal{L}+\sqrt{g}\delta_\xi\mathcal{L}=\sqrt{g}\nabla_\mu\left(\mathcal{L}\xi^\mu\right),
\end{align}
after using the fact that $\delta_\xi\sqrt{g}=\frac12\sqrt{g}g^{\alpha\beta}\delta_{\xi}g_{\alpha\beta}$. For the right hand side of \eqref{eq:genBact} we obtain:
\begin{align}\label{eq:Rhsga}
\begin{split}
\delta_\xi\left(\sqrt{g}\nabla_\mu B^\mu\right)=&\delta_\xi\left(\sqrt{g}\frac{1}{\sqrt{g}}\partial_\mu \left(\sqrt{g}B^\mu\right)\right)\\
=&\partial_\mu\left[\delta_\xi\left(\sqrt{g}B^\mu\right)\right]=\sqrt{g}\nabla_\mu\left[\nabla_\alpha\left(B^\mu\xi^\alpha\right)-B^\alpha\nabla_\alpha\xi^\mu\right].
\end{split}
\end{align}
Equating the results of \eqref{eq:Lhsga} and \eqref{eq:Rhsga} gives:
\begin{align}
\begin{split}
\sqrt{g}\nabla_\mu\left(\mathcal{L}\xi^\mu\right)=&\sqrt{g}\nabla_\mu\left[\nabla_\alpha\left(B^\mu\xi^\alpha\right)-B^\alpha\nabla_\alpha\xi^\mu\right]\\
\Rightarrow 0=&\nabla_\mu\left[\left(\mathcal{L}\xi^\mu\right)-\nabla_\alpha\left(B^\mu\xi^\alpha\right)+B^\alpha\nabla_\alpha\xi^\mu\right]=\nabla_\mu J^\mu
\end{split}
\end{align}
and thus, the N\"other current is given by:
\begin{align}\label{eq:BNC}
J^\mu\left[\xi\right]=\left(\mathcal{L}\xi^\mu\right)-\nabla_\alpha\left(B^\mu\xi^\alpha\right)+B^\alpha\nabla_\alpha\xi^\mu.
\end{align}
Referring back to \eqref{eq:genBact} we see that the N\"other current can also be written as the total divergence of a second rank tensor given by:
\begin{align}\label{eq:NC1and2}
\begin{split}
J^\mu\left[\xi\right]=\nabla_\alpha\left(B^\alpha\xi^\mu\right)-\nabla_\alpha\left(B^\mu\xi^\alpha\right)=\nabla_\alpha J^{\mu\alpha}\\
\Rightarrow J^{\mu\alpha}=B^\alpha\xi^\mu-B^\mu\xi^\alpha
\end{split}
\end{align}
From the above results we define the boundary N\"other charge in terms of \eqref{eq:BNC}:
\begin{align}\label{eq:NCr1F}
\mathcal{Q}[\xi]=\int_\Sigma d\Sigma_\mu J^\mu,
\end{align}
where $\Sigma$ is the timelike hyper surface with unit normal $m^\alpha$. Using Stoke's Theorem \cite{Poisson:2009pwt} we can rewrite $\mathcal{Q}[\xi]$ in terms of $J^{\alpha\beta}$ as:
\begin{align}\label{eq:NCr2F}
\mathcal{Q}[\xi]=\frac12\int_{\partial\Sigma} \sqrt{h}d\Sigma_{\alpha\beta} J^{\alpha\beta},
\end{align}
where $\sqrt{h}d\Sigma_{\alpha\beta}=-\sqrt{h}d^2x\left(n_\alpha m_\beta-n_\beta m_\alpha\right)$ is the area element of $\partial\Sigma$ and $n^\alpha$ is its spacelike  unit vector.

At this point we can demonstrate how the above formalism thus far relates to black hole entropy by applying it to a previously unexplored (within the respective formalism) example spacetime of general relativity. The specific spacetime we are interested in is the near horizon near extremal Kerr metric \cite{Bardeen:1999px,Amsel:2009ev} in the finite mass/temperature gauge \cite{Button:2013rfa,Guneratne:2016kib} given by the line element:
\begin{align}\label{eq:NHNEKsp}
\begin{split}
ds^2=\frac{1+\cos^2\theta}{2}\left(-\frac{r^2-2GM r-a^2}{\ell^2}dt^2+\frac{\ell^2}{r^2-2GM r-a^2}dr^2+\ell^2d\theta^2\right)+&\\
\frac{2\ell^2\sin^2\theta}{1+\cos^2\theta}\left(d\phi+\frac{r-2GM}{\ell^2}dt\right)^2,&
\end{split}
\end{align}
where $a$ is the angular momentum per unit mass parameter and $\ell^2=r_+^2+a^2$. This particular gauge is relevant for our analysis since it has non-zero Hawking temperature proportional in its usual way to its surface gravity $\kappa=f'\left(r_+\right)/2$, where $f(r)=\frac{r^2-2GM r-a^2}{\ell^2}$. Additionally, since $\kappa$ is non-zero, this particular gauge provides us with a non-zero finite time regulator for $\mathcal{Q}$ and specifically, the respective Hawking temperature $T_H=\kappa/2\pi$ coincides exactly with the non-extremal general Kerr black hole temperature in standard Boyer-Lindquist coordinates \cite{Button:2013rfa,Guneratne:2016kib}.

For Einstein's general relativity we can write the general surface term as \cite{Padmanabhan:2010zzb,Majhi:2012nq,Ashworth:1998uj}
\begin{align}\label{eq:GRgST}
B^\alpha=\frac{2Q\indices{^\alpha_\lambda^\sigma_\gamma}g^{\beta\gamma}\Gamma\indices{^\lambda_\beta_\sigma}}{16\pi G},~\text{where}~Q\indices{^\alpha_\lambda^\sigma_\gamma}=\frac12\left(\delta\indices{^\alpha_\lambda}\delta\indices{^\sigma_\gamma}-\delta\indices{^\alpha_\gamma}\delta\indices{^\sigma_\lambda}\right)
\end{align}
and $\Gamma\indices{^\lambda_\beta_\sigma}$ is the usual metric compatible Levi-Civita connection. Next, the spacetime in \eqref{eq:NHNEKsp} has the standard temporal Killing isometry $\xi_{(t)}=\partial_t$. For this particular isometry choice the charge $\mathcal{Q}[\xi_{(t)}]$ (called the N\"other energy) has the specific physical interpretation as the product of the entropy and temperature of $\partial\Sigma$ \cite{Ashworth:1998uj,Ashworth:2000mm,Padmanabhan:2012bs}. For the metric in \eqref{eq:NHNEKsp}, $\partial\Sigma$ corresponds to the black hole horizon and thus $\left.\mathcal{Q}[\xi_{(t)}]\right|_{r_+}$ should be equal to the product of $T_H$ and $S_{BH}$ of \eqref{eq:htaen}. Using \eqref{eq:NHNEKsp}, \eqref{eq:GRgST} and \eqref{eq:NCr2F} to compute the charge, we have:
\begin{align}
\begin{split}
B^r=&-\frac{2f'\left(r_+\right)}{16\pi G\left(1+\cos^2\theta\right)},~B^\theta=-\frac{8\cos^2\theta\cot\theta}{16\pi G\ell^2\left(1+\cos^2\theta\right)^2},~\text{and}~d\Sigma_{tr}=-\frac{1+\cos^2\theta}{2}\\
&\Rightarrow\left.\mathcal{Q}[\xi]\right|_{r_+}=\frac12\int_{\partial\Sigma} \sqrt{h}d\Sigma_{\alpha\beta} J^{\alpha\beta}=\frac{1}{32\pi G}\int\sqrt{h}d^2x4\kappa=\frac{A\kappa}{8\pi G},
\end{split}
\end{align}
where $A$ is the black hole horizon area, as aforementioned. Multiplying the above by the finite time regulator $\beta=1/T_H=2\pi/\kappa$ gives:
\begin{align}
\begin{split}
\mathcal{Q}[\xi]\beta=\frac{A\kappa}{8\pi G}\frac{2\pi}{\kappa}=\frac{A}{4 G}
\end{split}
\end{align}
and thus, reproducing the Bekenstein-Hawking entropy formula for the near horizon near extremal Kerr metric from the boundary N\"other current, within Einstein gravity.

The above thermodynamic analysis infers that the black hole entropy is encoded within the boundary term of the total action. We will now explore fundamentally how this inference can be established from a statistical counting of canonical quantum microstates via the general definition of the generator algebra:
\begin{align}
\left[\mathcal{Q}_1,\mathcal{Q}_2\right]=\frac12\left[\delta_{\xi_1}\mathcal{Q}\left[\xi_2\right]-\delta_{\xi_2}\mathcal{Q}\left[\xi_1\right]\right].
\end{align}
Next starting with \eqref{eq:NCr1F}, let us look at $\delta_{\xi_1}\mathcal{Q}\left[\xi_2\right]$:
\begin{align}
\begin{split}
\delta_{\xi_1}\mathcal{Q}\left[\xi_2\right]=&\int d\Sigma_\mu \delta_{\xi_1}\left(\sqrt{g}J^\mu\left[\xi_2\right]\right)\\
=&\int \sqrt{g}d\Sigma_\mu \left\{\nabla_\alpha\left(\xi_1^\alpha J^\mu\left[\xi_2\right]-\xi_1^\mu J^\alpha\left[\xi_2\right]\right)\right\}\\
=&\int_{\partial\Sigma} \sqrt{h}d\Sigma_{\mu\nu}\xi_1^\nu J^\mu\left[\xi_2\right],
\end{split}
\end{align}
which implies;
\begin{align}\label{eq:GenAlg}
\begin{split}
\left[\mathcal{Q}_1,\mathcal{Q}_2\right]=&\frac12\left(\int_{\partial\Sigma} \sqrt{h}d\Sigma_{\mu\nu}\xi_1^\nu J^\mu\left[\xi_2\right]-\int_{\partial\Sigma} \sqrt{h}d\Sigma_{\mu\nu}\xi_2^\nu J^\mu\left[\xi_1\right]\right)\\
=&\frac12\int_{\partial\Sigma} \sqrt{h}d\Sigma_{\mu\nu}\left(\xi_2^\mu J^\nu\left[\xi_1\right]-\xi_1^\mu J^\nu\left[\xi_2\right]\right).
\end{split}
\end{align}
\section{Horizon Microstates and CWG: Vacuum}\label{sec:CQMNHNEK}
In the next two sections we will compute the generator algebra of specific black hole solutions of CWG. A natural starting point is the near horizon near extremal Kerr metric in \eqref{eq:NHNEKsp}, since it is a vacuum solution of general relativity and thus, also a vacuum solution of CWG. To begin, we rewrite \eqref{eq:NHNEKsp} in terms of the usual general functions:
\begin{align}\label{eq:NHNEKgf}
\begin{split}
ds^2=\frac{1+\cos^2\theta}{2}\left(-f\left(r_++\rho\right)dt^2+\frac{1}{f\left(r_++\rho\right)}d\rho^2+\ell^2d\theta^2\right)+&\\
\frac{2\ell^2\sin^2\theta}{1+\cos^2\theta}\left(d\phi-A\left(r_++\rho\right)dt\right)^2,&
\end{split}
\end{align}
where we have employed a Bondi-like radial shift $r\to r_++\rho$ thus placing the horizon limit at $\rho=0$. Next, we imposed the following near horizon metric fall off conditions \cite{kerrcft}:
\begin{align}\label{eq:MetFOC}
\delta g_{\mu\nu}=\left(\begin{array}{cccc}\mathcal{O}\left(\rho\right) & \mathcal{O}\left(1\right) & \mathcal{O}\left(1\right) & \mathcal{O}\left(1\right) \\ & \mathcal{O}\left(1\right) & \mathcal{O}\left(1\right) & \mathcal{O}\left(\frac{1}{\rho}\right) \\ &  & \mathcal{O}\left(1\right) & \mathcal{O}\left(1\right) \\ &  &  & \mathcal{O}\left(1\right)\end{array}\right),
\end{align}
where we have indexed $\mu$ and $\nu$ over $\{t,\rho,\theta,\phi\}$. The above boundary conditions are preserved by a copy of the conformal group generated by the diffeomorphism: 
\begin{align}\label{eq:cgdiff}
\xi_{\epsilon}=\left(\epsilon(t)-\frac{\rho}{f\left(r_++\rho\right)}\epsilon'(t)\right)\partial_t-\rho\epsilon'(t)\partial_\rho,
\end{align}
which is evident when choosing $\epsilon(t)$ to be normalized circle diffeomorphisms; $\frac{e^{in\kappa t}}{\kappa}$ and computing the classical Lie bracket in the horizon limit $\rho=0$, which yields the centrally extended Virasoro algebra:
\begin{align}
i\left\{\xi_m,\xi_n\right\}=(m-n)\xi_{m+n}-\frac{m^3}{2\kappa}\delta_{m+n,0}\partial_t.
\end{align}

The boundary action for CWG is given by \cite{Li:2015mda,Grumiller:2013mxa,Lu:2012xu}:
\begin{align}\label{eq:cwgBT}
S_B=\frac{\alpha_c}{4\pi}\int d^3x\sqrt{\sigma}W^{\mu\nu\alpha\beta}n_\mu n_\nu\nabla_\alpha n_\beta,
\end{align}
where $n_\mu=\left\{0,\sqrt{\frac{1+\cos^2\theta}{2f(r)}},0,0\right\}$ is the previously defined spacelike unit normal. From the action \eqref{eq:cwgBT} we have the boundary Lagrangian density:
\begin{align}\label{eq:cwgBLd}
\mathcal{L}=\frac{\alpha}{4\pi}W^{\mu\nu\alpha\beta}n_\mu n_\nu\nabla_\alpha n_\beta,
\end{align}
from which we obtain $B^\alpha=n^\alpha\mathcal{L}$. Now substituting these results and \eqref{eq:cgdiff} into \eqref{eq:NC1and2} through \eqref{eq:NCr2F} and \eqref{eq:GenAlg}, where $m_\mu=\left\{\sqrt{\frac{\left(1+\cos^2\theta\right)f(r)}{2}},0,0\right\}$, we obtain in the horizon limit:
\begin{align}
\begin{split}
\mathcal{Q}=\frac{\alpha_c}{3\pi\ell^2}\int_{\partial\Sigma}\sqrt{h}d^2x&\left[\left(1+\cos ^2\theta\right) \left(-32 \ell ^4 \sin^2\theta A'(r_+)^2+\ell ^2 (\cos(2 \theta )+3)^2f''(r_+)\right.\right.\\
&\left.\left.+44 \cos (2 \theta )-\cos (4 \theta)+21\right)\right]/\left[(\cos (2 \theta )+3)^4\right]\\
&\left(\kappa\epsilon(t)-\frac{\epsilon'(t)}{2}\right)
\end{split}
\end{align}
and
\begin{align}
\begin{split}
\left[\mathcal{Q}_1,\mathcal{Q}_2\right]=\frac{\alpha_c}{3\pi\ell^2}\int_{\partial\Sigma}\sqrt{h}d^2x&\left[\left(1+\cos ^2\theta\right) \left(-32 \ell ^4 \sin^2\theta A'(r_+)^2+\ell ^2 (\cos(2 \theta )+3)^2f''(r_+)\right.\right.\\
&\left.\left.+44 \cos (2 \theta )-\cos (4 \theta)+21\right)\right]/\left[(\cos (2 \theta )+3)^4\right]\\
&\left[\kappa\left(\epsilon_1\epsilon'_2-\epsilon_2\epsilon'_1\right)-\frac{\epsilon_1\epsilon''_2-\epsilon_2\epsilon''_1}{2}+\frac{\epsilon'_1\epsilon''_2-\epsilon'_2\epsilon''_1}{4\kappa}\right].
\end{split}
\end{align}
Next, upon substituting $\epsilon(t)=\frac{e^{in\kappa t}}{\kappa}$ and performing the integration over $\sqrt{h}d^2x$ we have:
\begin{align}
\begin{split}
\mathcal{Q}_m=\frac{\alpha_c}{4\pi}\left[A'\left(r_+\right)+f''\left(r_+\right)\right]A\delta_{m,0}
\end{split}
\end{align}
and
\begin{align}
\begin{split}
\left[\mathcal{Q}_m,\mathcal{Q}_n\right]=\frac{\alpha_c}{4\pi}\left[A'\left(r_+\right)+f''\left(r_+\right)\right]\left[iA(m-n)\delta_{m+n,0}-im^3\frac{A}{2}\delta_{m+n,0}\right],
\end{split}
\end{align}
which we recognize as the quantum Virasoro algebra of the boundary degrees of freedom. Furthermore we can read off the central charge and lowest Virasoro mode:
\begin{align}
\begin{split}
\mathcal{Q}_0=&\frac{\alpha_c}{4\pi}\left[A'\left(r_+\right)+f''\left(r_+\right)\right]A\\
\frac{c}{12}=&\frac{\alpha_c}{4\pi}\left[A'\left(r_+\right)+f''\left(r_+\right)\right]\frac{A}{2},
\end{split}
\end{align}
which after employing Cardy's formula yields the black hole entropy of the near horizon near extremal Kerr black hole within CWG:
\begin{align}
S_{BH}=2\pi\sqrt{\frac{\mathcal{Q}_0c}{6}}=\frac{\alpha_c}{2}\left[A'\left(r_+\right)+f''\left(r_+\right)\right]A.
\end{align}
Comparing the above to the Wald entropy for CWG \eqref{eq:CWGWEnt} of the same spacetime \eqref{eq:NHNEKgf} we have:
\begin{align}
S=-\frac{\alpha_c}{8}\int W_{\mu\nu\alpha\beta}\epsilon^{\mu\nu}\epsilon^{\alpha\beta}d\Sigma=\frac{\alpha_c}{2}\left[A'\left(r_+\right)+f''\left(r_+\right)\right]A,
\end{align}
which precisely agrees with the previous result above for the near horizon near extremal Kerr black hole from counting the canonical quantum microstates via the boundary N\"other current method. 
\section{Horizon Microstates and CWG: non-Vacuum}\label{sec:CQM-Max}
Now we turn our attention to non-vacuum solutions of CWG, which provides us with an additional venue to explore more unique spacetimes. Specifically, we focus on the near horizon near extremal CWG-Maxwell theory. The CWG-Maxwell system has been well explored for the Coulomb potential in \cite{Mannheim:1990ya} with solution given by the line element \eqref{eq:SphSymBH}, where: 
\begin{align}\label{eq:cwgMKu1}
f(r)=w+\frac{u}{r}+\gamma r-kr^2,~w=1-3\beta\gamma,~u=-\beta(2-3\beta\gamma)-\frac{Q^2}{8\alpha\gamma},
\end{align}
and $Q$ is the black hole charge parameter for the $U(1)$ vector potential $A_{\mu}=\left\{Q/r,0,0,0\right\}$. For simplicity and without loss of generality, we can explore the extremal limit of \eqref{eq:cwgMKu1} in the mathematically simplified case where $\beta=\frac{3k+\gamma^2}{9k\gamma}$ and with extremal limit given by $Q=\sqrt{-\frac{8\alpha}{3}}$. In this scenario the metric \eqref{eq:cwgMKu1} takes the form:
\begin{align}
ds^2=\frac{\left(3rk-\gamma\right)^3}{27rk^2}dt^2-\frac{27rk^2}{\left(3rk-\gamma\right)^3}dr^2+r^2d\Omega^2
\end{align}
and looking at the near horizon, of the above in the standard way, where $r\to r\lambda+\frac{\gamma}{3k}$, $t\sim t/\lambda$ and further in the limit as $\lambda\to0$, we get $g_{tt}\to\infty$. This makes it seem naively unclear how to investigate the near horizon near extremal limit of the CWG-Maxwell theory, without killing off degrees of freedom or running into divergences in the metric.

However, it is well known \cite{Garousi:2009zx,Chen:2010bsa,Chen:2011gz,Chen:2009ht} that in the near horizon near extremal limit of charged black holes the $U(1)$ gauge potential becomes linear:
\begin{align}
A_\mu=\{Qr,0,0,0\}.
\end{align}
Additionally, the CWG-Maxwell system is stationary for the Euler-Lagrange equation: 
\begin{align}
-2\alpha_cW_{\mu\nu}+\frac12T_{\mu\nu}=0,
\end{align}
where $T_{\mu\nu}=F_{\mu\alpha}F\indices{_\nu^\alpha}-\frac14g_{\mu\nu}F^2$ is the Maxwell stress tensor and $W_{\mu\nu}=2\nabla_\alpha\nabla_\beta W\indices{_\mu^\alpha^\beta_\nu}+R_{\alpha_\beta}W\indices{_\mu^\alpha^\beta_\nu}$ is the Bach tensor, both of which stem from the inverse metric variation of the CWG-Maxwell theory. Consequently and employing the ansatz \eqref{eq:KKSolAnz}, we are free to forge for near horizon near extremal CWG-Maxwell solutions directly from the above field equation with linear $U(1)$ potential, to aid in our analysis. Our current investigation includes the two following general solutions:
\begin{align}\label{eq:sphscwgads2g}
\begin{split}
ds^2=&-B(r)dt^2+\frac{1}{B(r)}dr^2+\ell^2d\Omega^2\\
=&-\left(\frac{r^3 \left(2 \alpha \left(c_1^2 \ell^4-4\right)-3Q^2\ell^4\right)}{24\alpha c_2\ell^4}+\frac{r^2c_1}{2}+r c_2\right)dt^2\\
&+\left(\frac{r^3 \left(2 \alpha \left(c_1^2 \ell^4-4\right)-3Q^2\ell^4\right)}{24\alpha c_2\ell^4}+\frac{r^2c_1}{2}+r c_2\right)^{-1}dr^2+\ell^2d\Omega^2
\end{split}
\end{align}
and
\begin{align}\label{eq:cwgadsNo2s}
\begin{split}
ds_{c_3}^2=&\sin\theta\left(-f(r)dt^2+\frac{1}{f(r)}dr^2+\ell^2d\theta^2\right)+\ell^2\sin\theta d\phi^2\\
=&\sin\theta\left(-\frac{r^2+c_3r}{\ell^2}dt^2+\frac{\ell^2}{r^2+c_3r}dr^2+\ell^2d\theta^2\right)+\ell^2\sin\theta d\phi^2,
\end{split}
\end{align}
where the $c_i$'s are constants. The above solutions classify (general relativisticly) globally both as $Petrov~type~D$ and additionally as $Segre~type~[(11)(1,1)]~(IID-M2(11))$ and $[(11),Z\bar Z]~(II-M3(111))$ respectively. Without loss of generality and to ensure local $AdS_2\times S^2$ topology, we can impose that $c_1=\sqrt{\frac{3Q^2\ell^4+8\alpha_c}{2\ell^4\alpha_c}}$, which implies for \eqref{eq:sphscwgads2g} that:
\begin{align}\label{eq:sphscwgads2}
\begin{split}
ds_{c_2}^2=&-\left(r^2\sqrt{\frac{3Q^2\ell^4+8\alpha_c}{8\ell^4\alpha_c}}+rc_2\right)dt^2+\left(r^2\sqrt{\frac{3Q^2\ell^4+8\alpha_c}{8\ell^4\alpha_c}}+rc_2\right)^{-1}dr^2\\
&+\ell^2d\Omega^2.
\end{split}
\end{align}
We have included the subscripts $c_2$ and $c_3$ on the above line elements in \eqref{eq:cwgadsNo2s} and \eqref{eq:sphscwgads2} in order to distinguish them in subsequent discussion/analysis.
Next, we will compute the black hole entropy of the above two solutions via the previously outlined N\"other current method and again compare our results to the Wald entropy formula. 

It turns out that the above solutions exhibit the same (or even more relaxed) near horizon fall off conditions \eqref{eq:MetFOC} and thus near horizon preserving isometry given by \eqref{eq:cgdiff}. This can be more readily demonstrated, since the above solutions are spherically symmetric, via the Bondi-like transformation:
\begin{align}
dt=du+\frac{1}{f\left(r_++\rho\right)}d\rho
\end{align}
and solving the Killing equation $\mathcal{L}_\xi g_{\mu\nu}=0$, which yields for both $ds^2_{c_2}$ and $ds^2_{c_3}$ cases: 
\begin{align}
\xi^u=&F(u,\theta,\phi)~\text{and}\\
\xi^\rho=&-\rho\partial_u F(u,\theta,\phi).
\end{align}
Transforming back to the temporal coordinate $t$, we have:
\begin{align}
\xi^t=&\epsilon(t,\theta,\phi)-\frac{\rho}{f\left(r_++\rho\right)}\partial_t\epsilon(t,\theta,\phi)~\text{and}\\
\xi^\rho=&-\rho\partial_t\epsilon(t,\theta,\phi),
\end{align}
which matches \eqref{eq:cgdiff} for an $\epsilon$ that exhibits angular dependence. Now, for both $ds^2_{c_2}$ and $ds^2_{c_3}$ cases we may chose $n_\mu=\left\{0,\frac{1}{\sqrt{f(r)}},0,0\right\}$,  $m_\mu=\left\{-\sqrt{f(r)},0,0,0\right\}$, $\epsilon(t,\theta,\phi)=\epsilon(t)$ and following the the same procedures from Sec.~\ref{sec:CQMNHNEK}, we find: 
\begin{align}
\mathcal{Q}=
\begin{cases}
\frac{\alpha_c}{4\pi}\frac{2-\ell^2f''\left(r_+\right)}{6\ell^2}\int_{\partial\Sigma}\sqrt{h}d^2x\left(\kappa\epsilon(t)-\frac{\epsilon'(t)}{2}\right)&\text{case}~c_2\\
\frac{\alpha_c}{4\pi}\frac{-f''\left(r_+\right)}{6}\int_{\partial\Sigma}\sqrt{h}d^2x\left(\kappa\epsilon(t)-\frac{\epsilon'(t)}{2}\right)&\text{case}~c_3
\end{cases}
\end{align}
and 
\begin{align}
\begin{split}
&\left[\mathcal{Q}_1,\mathcal{Q}_2\right]=\\
&\begin{cases}
\frac{\alpha_c}{4\pi}\frac{2-\ell^2f''\left(r_+\right)}{6\ell^2}\int_{\partial\Sigma}\sqrt{h}d^2x\left[\kappa\left(\epsilon_1\epsilon'_2-\epsilon_2\epsilon'_1\right)-\frac{\epsilon_1\epsilon''_2-\epsilon_2\epsilon''_1}{2}+\frac{\epsilon'_1\epsilon''_2-\epsilon'_2\epsilon''_1}{4\kappa}\right]&\text{case}~c_2\\
\frac{\alpha_c}{4\pi}\frac{-f''\left(r_+\right)}{6}\int_{\partial\Sigma}\sqrt{h}d^2x\left[\kappa\left(\epsilon_1\epsilon'_2-\epsilon_2\epsilon'_1\right)-\frac{\epsilon_1\epsilon''_2-\epsilon_2\epsilon''_1}{2}+\frac{\epsilon'_1\epsilon''_2-\epsilon'_2\epsilon''_1}{4\kappa}\right]&\text{case}~c_3
\end{cases}.
\end{split}
\end{align}
Now, substituting $\epsilon(t)=\frac{e^{in\kappa t}}{\kappa}$ and performing the integration over $\sqrt{h}d^2x$ we obtain:
\begin{align}
\mathcal{Q}_m=
\begin{cases}
\frac{\alpha_c}{4\pi}\frac{2-\ell^2f''\left(r_+\right)}{6\ell^2}A\delta_{m,0}&\text{case}~c_2\\
\frac{\alpha_c}{4\pi}\frac{-f''\left(r_+\right)}{6}A\delta_{m,0}&\text{case}~c_3
\end{cases}
\end{align}
and
\begin{align}
\left[\mathcal{Q}_m,\mathcal{Q}_n\right]=
\begin{cases}
\frac{\alpha_c}{4\pi}\frac{2-\ell^2f''\left(r_+\right)}{6\ell^2}\left[iA(m-n)\delta_{m+n,0}-im^3\frac{A}{2}\delta_{m+n,0}\right]&\text{case}~c_2\\
\frac{\alpha_c}{4\pi}\frac{-f''\left(r_+\right)}{6}\left[iA(m-n)\delta_{m+n,0}-im^3\frac{A}{2}\delta_{m+n,0}\right]&\text{case}~c_3
\end{cases},
\end{align}
which we again recognize as the quantum Virasoro algebra of the boundary degrees of freedom. Again, reading off the central charge and lowest Virasoro mode, we have:
\begin{align}
\begin{split}
\mathcal{Q}_0=
&\begin{cases}
\frac{\alpha_c}{4\pi}\frac{2-\ell^2f''\left(r_+\right)}{6\ell^2}A&\text{case}~c_2\\
\frac{\alpha_c}{4\pi}\frac{-f''\left(r_+\right)}{6}A&\text{case}~c_3
\end{cases}\\
\frac{c}{12}=
&\begin{cases}
\frac{\alpha_c}{4\pi}\frac{2-\ell^2f''\left(r_+\right)}{6\ell^2}\frac{A}{2}&\text{case}~c_2\\
\frac{\alpha_c}{4\pi}\frac{-f''\left(r_+\right)}{6}\frac{A}{2}&\text{case}~c_3
\end{cases}
\end{split}
\end{align}
and after employing Cardy's formula, we obtain the black hole entropy for $ds^2_{c_2}$ and $ds^2_{c_3}$ within CWG:
\begin{align}\label{eq:CsEntCF}
S_{BH}=2\pi\sqrt{\frac{\mathcal{Q}_0c}{6}}=
\begin{cases}
\frac{\alpha_c\left(2-\ell^2f''\left(r_+\right)\right)}{12\ell^2}A&\text{case}~c_2\\
\frac{\alpha_cf''\left(r_+\right)}{12}A&\text{case}~c_3
\end{cases}.
\end{align}
Finally, comparing the above results to the CWG Wald entropy formula \eqref{eq:CWGWEnt} we have:
\begin{align}
S=-\frac{\alpha_c}{8}\int W_{\mu\nu\alpha\beta}\epsilon^{\mu\nu}\epsilon^{\alpha\beta}d\Sigma=
\begin{cases}
\frac{\alpha_c\left(2-\ell^2f''\left(r_+\right)\right)}{12\ell^2}A&\text{case}~c_2\\
-\frac{\alpha_cf''\left(r_+\right)}{12}A&\text{case}~c_3
\end{cases},
\end{align}
which again precisely agrees with the results for $ds^2_{c_2}$ and $ds^2_{c_3}$ from counting the canonical quantum microstates via the boundary N\"other current method, modulo the minus sign in the $ds^2_{c_3}$ case. We should note that, in the $ds^2_{c_3}$ case, we could have chosen the minus sign after taking the square root in the above Cardy formula. 
\section{Quantum Fields and $AdS_2/CFT_1$ in The Background of CWG}\label{sec:QFieldsICST}
In this section we will draw upon semiclassical methods in order to explore the construction of a near horizon $CFT$ action which holographically encodes the black hole thermodynamics of the respective black hole choice. We will be focusing our efforts on the $ds^2_{c_2}$ case and following methods outlined in Ref.~\refcite{Button:2010kg,Button:2013rfa,ry,Majhi:2012st}. In order to extract the two dimensional field content relevant to our construction we will probe the spacetime with a minimally coupled scalar, then taking the near horizon limit and integrating out angular degrees of freedom via a Robinson-Wilczek two dimensional decomposition \cite{robwill}. The resulting two dimensional theory will exhibit a known quantum effective action akin to the Polyakov action: 
\begin{align}
\label{eq:qseffa}
S_{eff}\sim\bar\beta^\psi\int d^2x\sqrt{-g^{(2)}}R^{(2)}\frac{1}{\square_{g^{(2)}}}R^{(2)}+\cdots,
\end{align}
where $\bar\beta^\psi$ is proportional to the Weyl anomaly coefficient~\cite{tseytlin89,polyak}. Our next task will be to construct the respective effective theory and value of $\bar\beta^\psi$ to $s$-wave approximation within the CWG paradigm. We will also need to use our results from Sec.~\ref{sec:CQM-Max} in order to tune $\bar\beta^\psi$ to reproduce the respective black hole entropy of the $ds^2_{c_2}$ spacetime. The $s$-wave approximation is appropriate if we infer that our scalar probe has gravitational origins, in other words $ds^2_{c_2}$ exhibits a Kaluza-Klein decomposition:
\begin{align}
\label{eq:wrsskk}
\begin{split}
ds^2=~&-f(r)dt^2+f(r)^{-1}dr^2+\ell^2 e^{-2\psi(r)}\left[d\theta^2+\sin^2\theta \left(d\phi-Adt\right)^2\right]\\
 =~&ds^2_{2D}+\ell^2 e^{-2\psi(r)}\left[d\theta^2+\sin^2\theta \left(d\phi-A_\mu dx^\mu\right)^2\right],
\end{split}
\end{align}
where we have introduced a two-dimensional black hole coupled to a two-dimensional real scalar and two-dimensional $U(1)$ gauge field. In the case of $ds^2_{c_2}$, we see that the only allowable gauge field couplings are linear phase shifts given by $\phi\to\phi-At$, which are trivial. However, for spacetimes not exhibiting global spherical symmetry, the $s$-wave is still an appropriate approximation, since the important gravitational dynamics seem to be contained in this regime \cite{strom1}.

We will begin our construction by extracting the contributing two-dimensional near-horizon field content by considering the four dimensional free scalar field in the background of \eqref{eq:wrsskk}:
\begin{align}
\label{eq:freescalar4}
S^{(4)}[\varphi,g]=&-\frac12\int d^4x\sqrt{-g}\nabla_{\mu}\varphi\nabla^\mu\varphi.
\end{align}
Next, we expand $\varphi$ in terms of spherical harmonics and examine the regime where $r\sim r_+$ in tortoise coordinates. Then, integrating out angular degrees of freedom, we are left with:
\begin{align}
\label{eq:tdtrw}
S^{(4)}[\varphi,g]\overset{r\sim r_+}{\longrightarrow}S^{(2)}[\varphi_{lm},g^{(2)}]=
\frac{\ell^2}{2}\int dtdr\varphi^*_{lm}\left[\frac{1}{f(r)}\left(\partial t-iA_t\right)^2-\partial_rf(r)\partial_r\right]\varphi_{lm}.
\end{align}
This simplified action is the result of the fact that in the examined regime, interacting/mixing and potential terms $(\sim l(l+1)\ldots)$ are weighted by the near horizion exponentially decaying term $f\left(r\left(r^*\right)\right)\sim e^{2\kappa r^*}$. Thus, in the respective regime, \eqref{eq:freescalar4} simplifies to an action of infinite massless complex scalar fields given by:
\begin{align}
\label{eq:nhwrss}
S=-\frac{\ell^2}{2}\int d^2x\varphi^{*}_{lm}D_{\mu}\left[\sqrt{-g^{(2)}}g^{\mu\nu}_{(2)}D_{\nu}\right]\varphi_{lm},
\end{align}
where $D_\mu=\partial_\mu-imA_\mu$ is the gauge covariant derivative. We have now arrived at the Robinson and Wilczek two-dimensional analog (RW2DA) fields for the \eqref{eq:wrsskk} spacetime represented by:
\begin{align}
\label{eq:2drwamet}
\g{^{(2)}_\mu_\nu}=~&\left(\begin{array}{cc}-f(r) & 0 \\0 & \frac{1}{f(r)}\end{array}\right)&f(r)=r^2\sqrt{\frac{3Q^2\ell^4+8\alpha_c}{8\ell^4\alpha_c}}+rc_2
\end{align}
and trivial $U(1)$ gauge field given by:
\begin{align}
\label{eq:rw2dgf}
A=A_tdt=constant=0.
\end{align}
Now, taking the $s$-wave approximation and implementing the CWG inspired redefinition $\varphi_{00}=\sqrt{\gamma}\psi$ (where $\psi$ is unitless and $\gamma$ will be proportional to $\alpha_c$), the effective action for the above two dimensional theory \eqref{eq:nhwrss} will be given by the two parts \cite{Leutwyler:1984nd,isowill}:
\begin{align}\label{eq:nh2pcft}
\Gamma=&\Gamma_{grav}+\Gamma_{U(1)},
\end{align}
where
\begin{align}
\label{eq:twoeffac}
\begin{split}
\Gamma_{grav}=&\frac{\gamma\ell^2}{96\pi}\int d^2x\sqrt{-g^{(2)}}R^{(2)}\frac{1}{\square_{g^{(2)}}}R^{(2)}~\text{and}\\
\Gamma_{U(1)}=&\frac{\gamma e^2 \ell^2}{2\pi }\int F\frac{1}{\square_{g^{(2)}}}F.
\end{split}
\end{align}
Comparing the above to \eqr{eq:qseffa}, we see that in our CWG inspired case $\bar\beta^\psi=\frac{\gamma\ell^2}{96\pi}$. We should note that $\Gamma_{U(1)}=0$ in \eqref{eq:nh2pcft} above for the $ds^2_{c_2}$ case, however we will keep track of it until now for completeness. 

The next step will be to restore locality in the quantum effective action \eqref{eq:nh2pcft}. This may be done via the introduction of auxiliary scalar fields $\Phi$ and $B$, where:
\begin{align}\label{eq:afeqm}
\square_{g^{(2)}} \Phi=R^{(2)}~\mbox{and}~\square_{g^{(2)}} B=\epsilon^{\mu\nu}\partial_\mu A_\nu.
\end{align}
For general $f(r)$ and $A_t(r)$ the above reduces to the following differential equations:
\begin{align}
\label{eq:afsol1.1}
\begin{split}
-\frac{1}{f(r)}\partial_t^2\Phi+\partial_rf(r)\partial_r\Phi=&R^{(2)}\\
-\frac{1}{f(r)}\partial_t^2B+\partial_rf(r)\partial_rB=&-\partial_r A_t,
\end{split}
\end{align}
with solutions given by:
\begin{align}
\label{eq:afsol}
\begin{split}
\Phi(t,r)=&\alpha_1 t+\int dr\frac{\alpha_2-f'(r)}{f(r)}\\
B(t,r)=&\beta_1 t+\int dr\frac{\beta_2-A_t(r)}{f(r)},
\end{split}
\end{align}
where $\alpha_i$ and $\beta_i$ are integration constants. Now, substituting \eqref{eq:afeqm} into \eqref{eq:twoeffac} yields our final near horizon $CFT$ in Liouville form given by:
\begin{align}\label{eq:nhlcft}
\begin{split}
S_{NHCFT}=&\frac{\gamma\ell^2}{96\pi}\int d^2x\sqrt{-g^{(2)}}\left\{-\Phi\square_{g^{(2)}}\Phi+2\Phi R^{(2)}\right\}\\
&+\frac{\gamma e^2 \ell^2}{2\pi }\int d^2x\sqrt{-g^{(2)}}\left\{-B\square_{g^{(2)}}B+2B \left(\frac{\epsilon^{\mu\nu}}{\sqrt{-g^{(2)}}}\right)\partial_\mu A_\nu\right\}
\end{split}
\end{align}

For the next part, we investigate and compute the non trivial asymptotic symmetries for the two dimensional gravitational part of \eqr{eq:2drwamet} for large $r$ behavior given by:
\begin{align}\label{eq:wrss2dwa}
\g{^{(0)}_\mu_\nu}=&
\left(
\begin{array}{cc}
-\frac{\sqrt{\frac{3Q^2\ell^4+8\alpha_c}{\ell^4 \alpha_c}}r^2}{2\sqrt{2}}-c_2r+\mathcal{O}\left(\left(\frac{1}{r}\right)^3\right)& 0 \\
 0 & \frac{2\sqrt{2}}{\sqrt{\frac{3Q^2\ell^4+8\alpha_c}{\ell^4 \alpha_c}}r^2}+\mathcal{O}\left(\left(\frac{1}{r}\right)^3\right) 
\end{array}
\right),
\end{align}
which is asymptotically $AdS_2$ with Ricci Scalar, $R=-\frac{2}{l^2}+O\left(\left(\frac{1}{r}\right)^1\right)$, where $l^2=\frac{2\sqrt{2}}{\sqrt{\frac{3Q^2\ell^4+8\alpha_c}{\ell^4\alpha_c}}}$. Coupling the above with the following metric fall-off conditions:
\begin{align}\label{eq:mbc}
\delta g_{\mu\nu}=
\left(
\begin{array}{cc}
    \mathcal{O}\left(r\right)&
   \mathcal{O}\left(\left(\frac{1}{r}\right)^0\right) \\
 \mathcal{O}\left(\left(\frac{1}{r}\right)^0\right) &
 \mathcal{O}\left(\left(\frac{1}{r}\right)^3\right)
\end{array}
\right),
\end{align}
we obtain the following set of asymptotic symmetries:
\begin{align}\label{eq:dpr}
\chi=-C_1\frac{r\sqrt{\frac{16}{\ell^4}+\frac{6Q^2}{\alpha_c}} \xi(t)}{4c_2+r\sqrt{\frac{16}{\ell^4}+\frac{6Q^2}{\alpha_c}}}\partial_t+C_2r\xi'(t)\partial_r,
\end{align}
for arbitrary function $\xi(t)$ and appropriate normalization constants $C_i$. Next, transforming to conformal light cone coordinates, which are given by:
\begin{align}
x^\pm=t\pm r^*,
\end{align}
where $r^*$ is the radial tortoise coordinate; we obtain:
\begin{align}
\label{eq:lcdiff}
\chi^\pm=\frac{\left(-C_1r\left(r^*\right) \sqrt{\frac{16}{\ell^4}+\frac{6Q^2}{\alpha_c}} \xi(x^+,x^-)\pm 4C_2 \xi'(x^+,x^-)\right)}{4c_2+r\left(r^*\right)\sqrt{\frac{16}{\ell^4}+\frac{6Q^2}{\alpha_c}}}.
\end{align}
The above diffeomorphisms are smooth on the respective radial boundary and the arbitrary constants can be normalized in order for $\chi^\pm$ to satisfy a Witt or $Diff$$\left(S^1\right)$ subalgebra $i\left\{\chi^\pm_m,\chi^\pm_n\right\}=(m-n)\chi^\pm_{(m+n)}$ when choosing a circle diffeomorphism representation for $\xi(x^+,x^-)$. 

The energy-momentum tensor for \eqref{eq:nhlcft} is given by the standard definition:
\begin{align}
\label{eq:emt}
\begin{split}
\left\langle T_{\mu\nu}\right\rangle=&\frac{2}{\sqrt{-g^{(2)}}}\frac{\delta S_{NHCFT}}{\delta g\indices{^{(2)}^\mu^\nu}}\\
=&\frac{\gamma\ell^2}{48\pi G}\left\{\partial_\mu\Phi\partial_\nu\Phi-2\nabla_\mu\partial_\nu\Phi+g\indices{^{(2)}_\mu_\nu}\left[2R^{(2)}-\frac12\nabla_\alpha\Phi\nabla^\alpha\Phi\right]\right\}.
\end{split}
\end{align}
At this point it will be advantageous to pick a convenient choice for the $c_2$ constant in \eqref{eq:2drwamet}, such that: 
\begin{align}
\begin{split}
f(r)=&\frac{r^2-\ell r}{l^2},~\text{where}\\
l^2=&\sqrt{\frac{8\ell^4\alpha_c}{3Q^2\ell^4+8\alpha_c}},~\text{as previously defined and}\\
c_2=& -\frac{\ell}{l^2}.
\end{split}
\end{align}
Now, using the general solution \eqr{eq:afsol} and substituting into \eqr{eq:emt}, and additionally implementing modified Unruh Vacuum boundary conditions (MUBC) \cite{unruh}:
\begin{align}
\label{eq:ubc}
\begin{cases}
\left\langle T_{++}\right\rangle=0&r\rightarrow\infty,~l\rightarrow\infty\\
\left\langle T_{--}\right\rangle=0&r\rightarrow r_+
\end{cases}
\end{align}
we are able to determine the integration constants $\alpha_i$:
\begin{align}
\begin{split}
\alpha_1=&-\alpha_2=\frac12f'(r_+)
\end{split}
\end{align}
and thus, the EMT. Given the two-dimensional reduction, our above EMT exhibits a Weyl (trace) anomaly, which is given by:
\begin{align}
\label{eq:tra}
\left\langle T\indices{_\mu^\mu}\right\rangle=-\bar\beta^\psi{4}R^{(2)},
\end{align}
which uniquely determines the value of central charge via \cite{cft}:
\begin{align}
\label{eq:center1}
\frac{c}{24\pi}=4\bar\beta^\psi\Rightarrow c=\ell^2\gamma=\frac{\gamma}{4\pi}A.
\end{align}
In addition and due to our use of the previously defined MUBC, the specific asymptotic boundary to $\mathcal{O}(\frac{1}{\ell})^2$ (which we denote by the single limit $x^+\to\infty$), implies that the EMT is dominated by one holomorphic component $\left\langle T_{--}\right\rangle$. When expanding this respective component in terms of boundary fields \eqr{eq:wrss2dwa} and computing the response to the respective asymptotic symmetries, we obtain:
\begin{align}
\delta_{\chi^-}\left\langle T_{--}\right\rangle=\chi^-\left\langle T_{--}\right\rangle'+2\left\langle T_{--}\right\rangle\left(\chi^-\right)'+\frac{c}{24\pi}\left(\chi^-\right)'''+\mathcal{O}\left(\left(\frac{1}{r}\right)^3\right),
\end{align}
where the primes indicate time derivatives. Thus, the above result tells us that $\left\langle T_{--}\right\rangle$ transforms asymptotically as the EMT of a $CFT$ of one dimension, with center given by \eqref{eq:center1}.

Given the above, we are now able to compute the generator charge alegra of the full asymptotic symmetry group, by compactifying (which regulates the asymptotic charges) the $x^-$ coordinate to a circle parametrized from $0\to 2\pi/\kappa$, where we define the respective asymptotic conserved charge as:
\begin{align}
\label{eq:asycharge}
\mathcal{Q}_n=\lim_{x^+\to\infty}\int dx^\mu\left\langle T_{\mu\nu}\right\rangle\chi^\nu_n.
\end{align}
Next, we replace $\xi(x^+,x^-)$ with circle diffieomorphisms $\frac{e^{-in\kappa x^\pm}}{\kappa}$ in \eqr{eq:lcdiff} and calculate the canonical response of $\mathcal{Q}_n$ with respect to the asymptotic symmetry, yielding:
\begin{align}
\label{eq:ca}
\delta_{\chi^-_m}\Q_n=\left[\Q_m,\Q_n\right]=(m-n)\Q_n+\frac{c}{12}m\left(m^2-1\right)\delta_{m+n,0}.
\end{align}
In other words, the asymptotic quantum generator algebra is given by a centrally extended Virasoro algebra, with center given by \eqr{eq:center1} and with lowest eigen-mode given by: 
\begin{align}
\label{eq:q0}
\Q_0=\frac{\gamma}{96\pi}A.
\end{align}

What we have shown thus far, is that the RW2DA fields of $ds^2_{c_2}$ given in \eqref{eq:2drwamet} are essentially descried by a one dimensional string theory given by \eqref{eq:nhlcft} and whose horizon microstates are holographically dual to a $CFT_1$ with center given by:
\begin{align}\label{eq:cre}
c&=\frac{\gamma}{4\pi}A
\end{align}
and lowest Virasoro eigen-mode:
\begin{align}
\Q_0&=\frac{\gamma}{96\pi}A.
\end{align}
Using these results in the Cardy Formula \eqr{eq:cf} we obtain:
\begin{align}
S=2\pi\sqrt{\frac{c\Q_0}{6}}=\frac{\gamma A}{24}.
\end{align}
Comparing the above to \eqref{eq:CsEntCF} allows us to fix the $\gamma$ constant for the $ds^2_{c_2}$ case, which yields:
\begin{align}
\gamma=\frac{8\pi\alpha_c\left(2-\ell^2f''\left(r_+\right)\right)}{\ell^2}.
\end{align}

Next, and now that we have fixed $\gamma$ and have computed the center \eqref{eq:center1} we can show how our specific $AdS/CFT$ construction also encodes black hole temperature by focusing on the EMT, which on the horizon is determined by one holomorphic component:
\begin{align}
\label{eq:hhf}
\left\langle T_{++}\right\rangle=-\frac{c}{48 \pi }\left(\frac{f\left(r_+\right)}{2}\right)^2.
\end{align}
The above is precisely the Hawking flux $(HF)$ modulo the central charge:
\begin{align}
\left\langle T_{++}\right\rangle=cHF=-\frac{c\pi}{12}\left(T_H\right)^2,
\end{align}
where $T_H=\frac{f\left(r_+\right)}{4\pi}$. These are interesting results, in that it demonstrates that our constructed $AdS_2/CFT_1$ correspondence contains both black hole entropy and black hole temperature.
\section{Discussion and Concluding Remarks}
In this paper we explored thermodynamic properties of black hole solutions within the CWG paradigm. We examined both vacuum and non-vacuum solutions and computed the respective black hole entropy of several different solutions within the N\"other current method. We further corroborated our results from the N\"other current method by comparing them to the Wald entropy formula for CWG. Despite our results agreeing with Wald's entropy, we have further shed light on the non-universality of black hole entropy within CWG, depending on the specific black hole examined and their symmetries. This is obviously different compared to Einstein-Hilbert gravity, where the black hole entropy seems to be given by the standard Bekenstein-Hawking formula irrespective of spacetime. 

Our N\"other current analysis implied that the respective quantum black hole horizon degrees of freedom are dual to a conformal field theory of lesser dimension. Thus, in our last task, we investigated this duality in more detail by constructing a two dimensional near horizon quantum effective action via semiclassical methods. We then showed how the resulting EMT, at the appropriate boundary, forms one copy of a centrally extended Virasoro algebra whose center and lowest eigen-mode encode calc hole entropy via Cardy's formula. We also demonstrate how our $CFT$ construction also contains black hole temperature via the quantum holomorphic flux of the $CFT$ on the horizon. 

One interesting caveat in our $AdS/CFT$ construction is the presence of a parameter $\gamma$, which needed to be fixed by comparing the resulting entropy to that obtained from the N\"other current method. It would be interesting for future work to perform a renormalization group flow analysis on this parameter, since it appears as part of the two dimensional $CFT$'s coupling. This may provide new insight on an analog $c$-theorem \cite{ctzjet} for CWG.
\section*{Acknowledgments}
We thank Vincent Rodgers and Philip Mannheim for enlightening discussions.
\appendix
\bibliography{cftgr}
\bibliographystyle{ws-ijmpd}
\end{document}